\documentstyle[aps,preprint,eqsecnum,epsfig]{revtex}
\begin{document}
\draft
\author{Joseph Rudnick and William Lay}
\address{Department of Physics and Astronomy, UCLA, 405 Hilgard Ave.,\\
 Los Angeles,  California 90095-1547}
\author{David Jasnow}
\address{Department of Physics and Astronomy, University of Pittsburgh,
Pittsburgh, PA 15260}
\date{\today}
\title{The effective potential, critical point scaling and the
renormalization group}
\maketitle
\begin{abstract}
The desirability of evaluating the effective
potential   in field theories near a phase transition
has been recognized in a number of different areas. We
show that recent Monte Carlo simulations for the
probability distribution for  the order parameter in an
equilibrium Ising system, when  combined with
low-order renormalization group results for an
ordinary $\phi^4$ system, can be used to extract the
effective potential. All  scaling features are
included in the process.
\end{abstract}

\section{Introduction}
\label{sec:intro}
In the realm of statistical physics, as well as in quantum field theory,
the need for non-perturbative approaches to study the vicinity of
a phase transition has long been appreciated.
Results of recent simulations\cite{T>T_c,T<T_c,T=T_c,Dohm} show that
the form of the effective potential for a variety of systems can be
determined using Monte Carlo methods.  Such methods have also been
applied to the study of the electroweak phase transition via lattice
Monte Carlo methods~\cite{kajantie}.  Additional continuum limit work
has recently been performed on a $\lambda \phi^4$ theory in
three-dimensions~\cite{bimonte}.  Extracting the effective potential
provides one with an additional window onto the equilibrium -- and
possibly the dynamical -- behavior of a system near and at its
critical point.  One use of Monte Carlo methods is to test scaling forms
based on the Renormalization Group.  A second is to explore the
equilibrium behavior associated with such phenomena as two-phase
coexistence and broken continuous symmetry.

Binder~\cite{Binder} and more recently Tsypin~\cite{T>T_c} and
Chen and Dohm~\cite{chen} have  utilized Monte Carlo and binning
techniques to
generate probability distributions of the order parameter of an $O(1)$
(Ising-like) system in the vicinity of its critical point.  In Tsypin's work
data were generated at a single
temperature, and over a range of symmetry-breaking fields
in order to explore the full scaling domain. Above the critical
temperature  the distribution was found to be consistent with a low-order
polynomial form for the effective potential,
which, as discussed below, is essentially the logarithm of the probability
distribution.
The results of simulations were
most closely fit when the effective potential was terminated at the sixth
order. The standard quartic Ginzburg-Landau form for the effective potential
failed to produce an adequate fit for any choice of coefficients, and the
addition of eighth and higher order terms did not materially improve the
agreement between the fitting form and the results of simulations.

This remarkable set of results persists below the critical
temperature~\cite{T<T_c}. Here, it was found that a sixth-order polynomial form
for the effective potential leads to an outstanding match to the results of
simulations. In this case the best-fit effective potential was found to
contain a {\em negative} quartic term.
An additional feature of Tsypin's fitting form is a prefactor, proportional to
the square root of the second derivative of the effective potential.

It is reasonable to ask whether or not the data are consistent with
a scaling form for the
effective potential. In fact, the nature of the polynomial strongly suggests
a form incorporating thermodynamic scaling with exponents appropriate to
the Ising model. The rationale for this will be explored in more depth in
subsequent sections.

It should be noted that the renormalization group has been utilized
previously to generate order parameter distributions in the vicinity
of the critical point\cite{Dohm,chen,Bruce}.  The principal novelty in
this paper is the attempt to produce a unified form for the order
parameter distribution that fits data above, below and at the critical
point.  It is found that the inclusion of the prefactor mentioned
above leads for excellent agreement with data based on Monte Carlo
simulations.  The prefactor does not play an important role in the
fitting of a scaling form to the order parameter distribution in the
high and low temperature phases.  However, this finite size correction is
essential to the construction of a high-precision fit to data at the critical
point.

Checks on the quality of the fit include comparisons
of a universal ratio of moments of the distribution with
values in the literature determined by a variety of alternative
techniques, and
comparison between our optimum fit value of the
fourth order coupling $u$ with a variety
of other
determinations of that quantity.  These checks are very encouraging.  It
appears that a scaling form based on renormalized mean field theory
agrees to a high degree of accuracy with published results.  On this
basis, one has confidence that an equation of state constructed in a
similar manner from renormalized mean field theory will accurately describe
systems in the $O(1)$, or Ising model,
universality class.

The remainder of this paper is laid out as follows. In the next
section the effective potential is defined and a phenomenological scaling
description is presented.
In Sec.~\ref{sec:RG} a renormalization group derivation of the
scaling form is given, and in the following section the ``prefactor''
is discussed.
In Sec.~\ref{sec:finite} finite size
effects are addressed, while Sec.~\ref{sec:compare}   assesses the
success of the scaling form of the effective potential as a fit to
Tsypin's Monte Carlo
simulations.  Sec.~\ref{sec:conc} is devoted to concluding remarks.
An Appendix contains a discussion of the effect of the prefactor on an
important universal quantity.

\section{scaling considerations}
\label{sec:scaling}
We imagine a scalar (Ising-like) system described by order
parameter, $\phi$ and ordering field, $h$, in equilibrium at temperature
$T$.  If the system is described by Hamiltonian $H(\phi)$ the
constrained (reduced) free energy, $F(T,h;M)$, is given by
\begin{eqnarray}
Z & = &  \exp\left[-F(t,h;M)\right] = \int D\phi\  \delta\left(M -
I(\phi)\right)\exp\left[-H(\phi)/T\right]\: , \end{eqnarray}
where $I(\phi) \equiv \Omega^{-1}\int \phi\ d^dx$ is the average
order in configuration $\{\phi\}$,
 $\Omega$ represents the volume of the system, and $d$ is the
dimensionality. The effective potential
 ${\cal H}$, is usually defined as
\begin{equation}
Z  = \exp( -{\cal H} )
\label{eff-pot-def}
\end{equation}
which identifies it with the constrained free energy.
According to standard scaling notions, when the linear extent, $L$, of
the system is sufficiently great, the effective potential, ${\cal H}$, of
an $O(1)$ system will have the form
\begin{equation}
{\cal H}\left(t,h,M \right) = L^{d}\left|M\right|^{2d/(d-2+\eta)} {\cal
F}\left(t\left| M \right|^{-2/\nu \left( d-2+\eta \right)},h\left| M
\right|^{-\left(d+2-\eta \right)/\left( d - 2 + \eta \right)} \right).
\label{effp}
\end{equation}
The parameter $t$ is the reduced temperature, $t = (T- T_c)/T_c$,
where $T_c$ is the critical
temperature, and the ordering, or symmetry-breaking, field $h$
has been introduced above.  The quantity $L$ is a characteristic
linear extent of the system, and $d$ is its spatial dimensionality.  The
exponents $\nu$ and $\eta$ appear in the (unconstrained) two-point
correlation function $C\left(\vec{r}-\vec{r^{\prime}},t,h\right)$,
\begin{equation}
C\left(\vec{R},t,h\right) = \left|\vec{R}\right|^{-\left(d-2+\eta \right)}
{\cal C}\left(\left|\vec{R}\right|t^{\nu},ht^{-\nu\left(d+2-\eta\right)/2}
\right).
\label{corr}
\end{equation}
All thermodynamic exponents follow from the two correlation function exponents,
$\nu$ and $\eta$, given the standard scaling and hyperscaling
relations\cite{scaling}. When the spatial dimensionality is $d=3$,
the critical exponents are~\cite{ising-exps}
\begin{eqnarray}
\nu &=& 0.63 \pm 0.002 \label{nu}\\
\eta &=& 0.037 \pm0.001 \label{eta}
\end{eqnarray}

An immediate consequence of expression (\ref{effp})  is that the effective
potential at the critical point  ($t = h = 0$)  has the form
\begin{equation}
{\cal H}_{\rm critical} \propto L^{d}\left|M \right|^{2d/\left(d-2+
\eta \right)} = L^{d}\left|M\right|^{\approx 5.7} \ .  \label{Hcrit}
\end{equation}
It is reasonable to expect that, within the scaling regime,
at low enough reduced temperature and ordering
field, and for appropriate values of the order parameter (that is, if
the order parameter controls the decay of correlations) the effective
potential will behave in the same way.  If one attempts to approximate
the effective potential by a polynomial in the order parameter, $M$,
the ``best'' one will, most likely, not have terms that go beyond
sixth order.  Interestingly, this is precisely the result of Tsypin's
unbiased attempt to fit his simulation data to a polynomial effective
potential: A {\em fourth order}, Ginzburg-Landau form {\em fails} to
adequately represent the data, while the coefficients of terms {\em
beyond sixth order} are so small as to cast doubt on their appearance
in the true effective potential.  This result holds both above and
below the critical temperature---that is, whether the system is in the
disordered or the ordered state.

Now, the form (\ref{effp})
is expected to apply only when the order parameter $M$ lies
at -- or close to -- the value which minimizes the free energy
at fixed $t,h$. In
particular, there are known difficulties in applying the above scaling form
(\ref{effp}) to
the case of a system in the coexistence region,
which impinge directly on the
scaling regime. The equilibrium state within the coexistence regime
will generally involve one or more interfacial regions separating
two homogeneous thermodynamic phases.
Because of this, the
scaling form, which is hypothesized
for a {\em single} homogeneous phase,
does not necessarily do an adequate job of describing the
behavior
of a system in which two different phases coexist.  On the other hand, a
scaling form ought to predict with great accuracy the dependence
of thermodynamic functions on temperature
and ordering field as these fields
approach the coexistence line,
that is, at the phase boundary for two-phase coexistence.

There is one
more consequence of scaling that is worthy of note.  The scaling form
in Eq.  (\ref{effp}) together with correlation function scaling
implies the following form for the free energy\cite{PF,Stauffer}
\begin{eqnarray}
{\cal H} &=& \hat{\cal
G}\left(M\xi^{(d-2+\eta)/2},h\xi^{(d+2-\eta)/2},t\xi^{1/\nu},L\xi^{-1}\right)
\nonumber \\
&=& \hat G\left(M|t|^{-\nu\left(d-2+\eta\right)/2},
h|t|^{-\nu\left(d+2-\eta\right)/2}, L|t|^{\nu}, \frac{t}{|t|}\right) \ .
\label{ghat}
\end{eqnarray}
In the first line of Eq.  (\ref{ghat}) the quantity $\xi$ is the {\em
bulk} correlation length at $h=0$.

\section{renormalization-group based form for the effective potential}
\label{sec:RG}
The general form in Eq. (\ref{effp}) restricts, but does not specify, the
detailed dependence of the effective potential ${\cal H}$
on $t$ and $h$. Other considerations are required for an
explicit evaluation. One candidate form is based on the
Renormalization Group, particularly the field-theoretical expansion
in $\epsilon = 4 - d$, where, as above, $d$ is the spatial
dimensionality of the system.
We consider a standard $\phi^4$ Hamiltonian
\begin{equation}
\frac{-H}{T}= \int [ \frac{1}{2}\left((\nabla \phi\right)^2 + r \phi^2 )
+ u \phi^4 ]\ d^dx \ .
\label{eq:phi-4}
\end{equation}
At lowest non-trivial order in $\epsilon$, the effective potential
has the form~\cite{nelson-rudnick}
\begin{equation}
{\cal H}(M,t,h) = e^{-\ell^*d}\left[\frac{1}{2}te^{\left(1/\nu \right)\ell^{*}}
M^2e^{\left(d-2+\eta
\right)\ell^{*}} + \frac{u}{4}M^4e^{2\left(d-2+\eta\right)\ell^{*}} -
hMe^{\ell^*d} \right].
\label{Greffp}
\end{equation}
This approximate effective potential is at the level of a
{\em renormalized} Ginzburg-Landau free
energy, and is hence equivalent to
``renormalized mean-field theory''.
The renormalization, indeed the full effect of the Renormalization
Group trajectories, resides in the quantity $e^{\ell^*}$ which is
equivalent to the ``block spin'' size. This key quantity may be determined
via
\begin{equation}
te^{\left(1/\nu \right)\ell^{*}} + 3 u M^2 e^{\left( d- 2 + \eta
\right)\ell^*}
= 1.
\label{l*}
\end{equation}
The effective potential described by Eqs. (\ref{Greffp}) and (\ref{l*})
is fully consistent with the scaling hypotheses
embodied in Eq. (\ref{effp}). It is is completely determined
once the coupling constant $u$ has been set, along with ``metrical'' factors
associated with the scales of $h$ and $t$. The inclusion of higher
order, loop expansion contributions leads to modifications of
the effective
potential, so, on the
face of it, there is no reason to expect that (\ref{Greffp}) will be exact.
However, given that it is entirely consistent with critical point scaling,
that it embodies the full RG trajectories,
and that it represents the leading order contribution in a systematic
Renormalization Group expansion for the free energy,
it seems at least a plausible ``zeroth order'' candidate for
the critical effective potential. This optimism is bolstered
by other situations in which ``renormalized mean-field theory''
provides surprisingly good numerical results when applied in
three dimensions.~\cite{jasnow-review}

\section{the prefactor}
\label{sec:prefactor}
In this section we discuss the prefactor introduced by
Binder~\cite{binderagain} and used by Tsypin in fitting Ising
simulations~\cite{T>T_c,T<T_c}.  A variety of arguments have been
advanced for the existence of this term.  Here we introduce additional
justification for the prefactor.

Suppose we allow spatial variations of the order parameter
around the globally constrained value $M$ in the form
$M +\sigma(\vec{k})$. Note that $\langle\sigma(\vec{k})\rangle=0$.
Neglecting terms of greater than quadratic order in the $\sigma({\vec k})$'s,
the dependence of the effective Hamiltonian on the
$\sigma(\vec{k})$'s
at gaussian order  will be
\begin{equation} \frac{H_{\rm gauss}\left[\sigma(\vec{k})\right]}{T} =
\sum_{\vec{k} \neq 0} \left[
\frac{1}{2}\left|\vec{\nabla} \sigma(\vec{k})\right|^2 + \frac{1}{2}{\cal
H}^{\prime \prime}\sigma(\vec{k}) \sigma(-\vec{k}) \right]. \label{gauss}
 \end{equation}
The term ${\cal H}^{\prime \prime}$ is the second derivative with respect to
the order parameter $M$ of the effective potential ${\cal H}$ at renormalized
mean field level
discussed above. As noted the
sum excludes the $\vec{k}=0$ mode of the order parameter fluctuations.
Integration
over the $\sigma(\vec{k})$'s yields the following ``gaussian'' correction to
the effective potential
\begin{equation}
\sum_{\vec{k} \neq 0} \frac{1}{2} \ln\left(\frac{{\cal H}^{\prime \prime} +
k^2}{2 \pi} \right) \rightarrow \frac{\Omega}{\left(2 \pi \right)^d}\int d^dk
\frac{1}{2} \ln\left(\frac{{\cal H}^{\prime \prime} + k^2}{2 \pi} \right) -
\frac{1}{2} \ln \left(\frac{{\cal H}^{\prime \prime}}{2 \pi} \right) .
\label{gauss1}
\end{equation}
As above $\Omega$ is the spatial volume of the system. The last term
on the right hand side of Eq. (\ref{gauss1}) is the leading order difference
between the sum on the left hand side and its asymptotic limit as
the integral on the right.
This term can be thought of as a
finite size correction to the extensive result in the limit,
$\Omega \rightarrow \infty$.
A general feature of the prefactor that will prove to be of some use shortly is
its connection with the renormalized mean-field susceptibility, $\chi$.
It is straightforward
to verify that the second derivative of the effective potential with respect to
the order parameter is inversely related to the susceptibility, i.e.,
\begin{equation}
{\cal H}^{\prime \prime}(M) = \frac{1}{\chi(M)}\ .  \label{chidef}
\end{equation}

When the effective potential displays the effects of critical fluctuations,
the derivation above acquires modifications,
specifically the inclusion of counter-terms in a full
renormalization group treatment.
Most of the steps leading
to this finite size correction are as displayed above. The final result
preserves the relationship between the prefactor and the isothermal
susceptibility implied by the above equation. Specifically, the new prefactor
has the form
\begin{equation}
\sqrt{\frac{{\cal H}^{\prime \prime}}{2 \pi}} = \frac{1}{\sqrt{2 \pi \chi}}.
\label{newprefactor}
\end{equation}
At the level of renormalized mean-field theory, one can write for the
susceptibility
\begin{equation}
\chi = e^{\left(2-\eta\right)\ell^*} = e^{\gamma \ell^*/\nu},
\label{renormalizedchi}
\end{equation}
where $\gamma$ is the critical exponent for the isothermal
susceptibility ($\chi \propto \left|t\right|^{-\gamma}$).
As noted previously, the quantity $\ell^*$ is determined by Eq. (\ref{l*}).

There is a more general argument for the existence of the prefactor.
This argument is based on the ``infinitesimal'' momentum shell version
of the Renormalization Group used in  the calculation of the partition
function of the Ginzburg-Landau-Wilson model\cite{Rudnick}.  In this
approach, modes at the surface of a shrinking Brillouin zone are
integrated out.  The contributions to the free energy have a
gaussian-like form.  The key contribution to the net free energy is
\begin{equation}
\frac{1}{2}\sum_{k} \log\left[\Sigma_{k} + k^{2}\right]
\label{freeenergy2}
\end{equation}
In expression (\ref{freeenergy2}), the self energy term $\Sigma_{k}$
contains the effects of fluctuations whose wave vectors exceed $k$ in
magnitude.  If the uniform mode is singled out, the sum in
(\ref{freeenergy2}) is carried out over all non-zero $k$'s.  The one
term that does not contribute to the free energy is, thus, equal to
\begin{equation}
\frac{1}{2} \log \left[ \Sigma_{0}\right]
\label{missingterm}
\end{equation}
In this term the restriction on the momenta of the fluctuations
contributing to
$\Sigma_0$ no longer differentiates the quantity from
the standard self energy, $\Sigma$.  Standard arguments suffice to establish
the connection between the the susceptibility and the self energy.  If
the $k=0$ term is added to the sum in Eq.
(\ref{freeenergy2}), we end up with a sum over \underline{all} $k$ that
constitutes the leading order bulk contribution to the effective
potential.  It is then necessary to perform a subtraction leading to a
term in the exponent of the form $ 1/2 \log \left[
\Sigma_{0}\right] = \ln \sqrt{\chi_{T}}$.  That the ``subtracted'' term
appears in the exponent with a positive sign reflects the fact that
the partition function is the exponential of minus the free energy.

\section{finite size effects}
\label{sec:finite}
There are other finite size effects. For example, there is a limit on the
maximum possible size of a block spin in the
renormalization procedure.  This limit appears as a modification
of Eq. (\ref{l*}). A limit on the block spin size, and hence the quantity
$\ell^{*}$, is enforced if Eq.  (\ref{l*})
is modified so as to ensure
that the size of the block spin does not exceed the dimensions of the
system.  The new requirement on $\ell^{*}$ is
\begin{equation}
te^{\left(1/\nu \right)\ell^{*}} + 3 u M^2 e^{\left( d- 2 + \eta
\right)\ell^*} + \left(\frac{c}{L}\right)^{2} e^{2\ell^{*}}
= 1. \label{cutoff}
\end{equation}
The quantity $c$ is a number of
order unity.  In principle, the results of a calculation will not depend
on the specific value of $c$, as the value of $\ell^{*}$ is a detail
of the renormalization group procedure, which has no effect on the
final result\cite{nelson-rudnick}.  In practice, the results that
follow from the use of Eq.  (\ref{cutoff}) depend sensitively on the
choice of the parameter $c$.
In the work reported here,
this quantity thus acquires the status of a
fitting {\em metrical} parameter connecting the lattice size in the
Ising simulations to the length parameter $L$ in the renormalization
group approach.

\section{Comparison with data}
\label{sec:compare}
The effective potential (\ref{Greffp}) is now compared to the
results of the Ising model simulations performed by
Tsypin\cite{T>T_c,T<T_c,T=T_c}. Four data sets from Tsypin's simulations are
fit, one in the ordered phase ($\beta_{o}=0.2227$), one in the disordered
($\beta_{d}=0.22055$) and two at the
critical temperature $T=T_{c}$; ($\beta_{c}=0.22165$). The data in the
disordered phase are gathered in a Monte Carlo investigation of the order
parameter distribution of a system of $58^3$ Ising spins.
In the case of
the ordered phase, the simulations were performed on a
system of $74^3$ spins, while
lattices of $16^{3}$ and $32^{3}$ were used for the simulations
at criticality. The
data are fit with the renormalized mean-field approximation Eq. (\ref{Greffp})
including the prefactor. The fitting procedure includes the use of the
restriction on
$\ell^{*}$ obtained by solving the transcendental equation,
(\ref{cutoff}).  The overall normalization is set such that the area under
our curve matches that of Tsypin's histograms.

Our fitting procedure utilizes a grid-search minimization of the
$\chi^2(\vec \alpha)$ merit function.
\begin{equation}
	\chi^{2}(\vec \alpha) = \frac{1}{D}\sum_{i}^{}\frac{(y(x_{i};\vec
	\alpha)-d_{i})^{2}}{\sigma_{i}^{2}}
	\label{chisq}
\end{equation}
The parameters $\{\alpha_j\}$ for the
zeroth-order fit are $t$, $u$ and $c$ that appear in (\ref{Greffp})
and (\ref{cutoff}).
Technically, this expression is the {\em reduced} $\chi^2$ since we
divide by the number of degrees of freedom, $D$. $D$ is the
number of data bins less the number of free parameters.
The errors $\sigma_{i}$ were assumed to be of Gaussian order; i.e.
$\sigma_{i}^{2}=y(x_{i};\vec \alpha)$. This assumption is
conservative since the $\chi^2$ is less than the expected result, which
should be approximately unity.

As an initial effort to minimize $\chi^2$, a {\em blind fit} is performed
across
all the histograms of a given data set (e.g., in the disordered phase
the seven
histograms corresponding
to the seven different values of the external field
$h$). See Table \ref{table1}.
In this blind fit each bin of each
histogram is treated with the same
weight. This procedure certainly leads to an unbiased minimization of $\chi^2$,
but further fine-tuning is required as explained below.

It is important to note the similarity between the values of $u$ in both
temperature regimes.
In zeroth-order approximation we find $u_d=u_o=0.231$
for the disordered and ordered phases, respectively.
This is quite significant in that Landau-Ginzburg theory dictates that the
fourth-order coupling be the same everywhere. Also note that the
value of $t$ that best fits the data from the critical simulations is
indeed zero, as expected. We do observe one anomaly in the best fit
values of the reduced temperature in the ordered and disordered
phases, however. Namely, if we compare the ratio of the temperatures
we obtain
\begin{displaymath}
	\frac{t_{d}}{-t_{o}}=0.873
\end{displaymath}
while from Tsypin's simulation we have
\begin{displaymath}
\frac{t_{d}}{-t_{o}}=-\frac{\beta_{d}-\beta_{c}}{\beta_{o}-\beta_{c}}=1.05.
\end{displaymath}
This is unfortunate as we expect this ratio to be the same in the
simulations and our fit even though the temperatures themselves need
not be.

An additional comment needs to be made for the fitting of the ordered
phase data. Notice that in the $h$ = 0 histogram we have removed the data
lying to the left of the peak. Attempts to fit the data in that regime did not
produce satisfactory results. We attribute this failure to the fact that
a single
renormalized Ginzburg-Landau form (such as Eq.~(\ref{Greffp}))
is inappropriate to the region of two-phase
coexistence.

One expects that the scaling form for the free energy
(combined with the prefactor) should provide a reasonable
approximation to the free energy in {\em all} temperature regimes
using a single value of $u$ and $c$. To
this end, we search for a set of parameters that minimize the
$\chi^{2}$ for all data sets. The most effective strategy for
accomplishing this task is to first determine a range of parameters
$u$ and $c$ that fit the $t=0$ data. These
bounds are then used for a three parameter fit in the $t>0$ and $t<0$
phases which
yields final values of $t$, $u$ and $c$. It should be noted that a
significant amount of ``tweaking'' of the parameters is necessary in
addition to the blind minimization of the $\chi^{2}$.
These values are tabulated in \ref{table2} and the resulting
probability densities are plotted against Tsypin's data in the three
temperature regimes.  The plots are displayed in Figures 1 - 3.  It is
interesting to note that the ratio of temperatures discussed above is
now
\begin{displaymath}
	\frac{t_{d}}{-t_{o}}=1.06
\end{displaymath}
which is within one percent of Tsypin's value.

To further evaluate our scaling form we also calculate the universal
quantity\cite{B&K,Binder}
\begin{eqnarray}
\Gamma_4&=&\frac{\langle M^4 \rangle }{\left( \langle M^2 \rangle \right)^2}
- 3
\label{Gamma}
\end{eqnarray}
 The values of $\Gamma_4$ from our
calculations
compared to other sources are tabulated in \ref{table3}.

As a final check on our results, we compare our best fit value of the
renormalized coupling constant $u$ with results obtained by other
methods.  A variety of techniques have been utilized to determine
this universal quantity, including the $\epsilon$ expansion\cite{P&V},
high order loop expansions in three dimensions\cite{G&Z,M&N,A&S},
Monte Carlo simulations and high temperature series\cite{refs}.
Recently reported values of the renormalized fourth order coupling
constant range from 0.233\cite{M&N} to 0.236\cite{G&Z}.  This is
to be compared with our optimum global fit $u=0.228$; see Table
\ref{table2}. Given that no attempt twas made to match previously
determined values of the renormalized fourth order coupling in our
fitting procedure, the quality of agreement can be described as, at
the very least, encouraging.

 \begin{table}
{
\caption{The values of fitted parameters $t$, $u$ and $c$ obtained
from the blind fits for four different data sets.
The stated errors are 68\%\ confidence limits; i.e. one standard deviation.}
\label{table1}
\begin{tabular}{ccccc}
&{$T=T_c$} &{$T=T_c$} &{$T>T_c$} & {$T<T_c$}\\
   \tableline
    $L$      &$16$ &$32$ &$58$ &$74$\\
    $t$ 	 &$0\:^{+0.0002}_{-0.0002}$ &$0\:^{+0.00006}_{-0.00007}$
             & $0.00356^{+0.0002}_{-0.0003}$
             & $-0.00418^{+0.0003}_{-0.0003}$\\
     $u$     &$0.224^{+0.005}_{-0.007}$  &$0.217^{+0.005}_{-0.005}$
             & $0.231^{+0.004}_{-0.004}$ & $0.231^{+0.003}_{-0.003}$\\
     $c$     & $0.437\;^{+0.04}_{-0.03}$ & $0.458\;^{+0.03}_{-0.03}$
             & $1.79\;^{+1.2}_{-1.8}$
        \tablenote{$\:$Due to the extremely
        weak dependence of the free energy on the variable $c$ in the
$T\neq T_{c}$
        phases there is a large range of values  within the 68\% confidence
limit.}
             & $1.60\;^{+0.9}_{-1.6}$ \\
$\chi^2$        & $0.000335$ & $0.000237$  & $0.0170$  	& $0.0388$\\
  \end{tabular}
}
\end{table}

\begin{table}
{
\caption{The values of fitted parameters $t$, $u$ and $c$ after
constraining to fit all temperature regimes simultaneously.}
\label{table2}
\begin{tabular}{cccc}
&{$T=T_c$} &{$T>T_c$} & {$T<T_c$}\\
   \tableline
    $L$      &$16, 32$
    \tablenote{$\:$We fit both $L=16$ and $L=32$ data sets simultaneously.}
             &$58$ &$74$\\
    $t$ 	 &$0\:^{+0.0001}_{-0.0003}$
             & $0.00408^{+0.0003}_{-0.0006}$
             & $-0.00385^{+0.0002}_{-0.0004}$\\
     $u$     &$0.228^{+0.006}_{-0.023}$
             & $0.228^{+0.008}_{-0.005}$ & $0.228^{+0.006}_{-0.003}$\\
     $c$     & $0.45\;^{+0.10}_{-0.07}$
             & $0.45\;^{+2.5}_{-0.45}$ & $0.45\;^{+2.5}_{-0.45}$ \\
$\chi^2$        & $0.00112$ & $0.0366$ & $0.0703$\\
  \end{tabular}
}
\end{table}

\begin{table}
{
\caption{The value of the universal quantity $\Gamma_4$ calculated for $d=3$
Ising systems at $t=0$ from various sources.}  \label{table3}
\begin{tabular}{ccc}
    source		   	&$L$    &$\Gamma_4$   \\
   \tableline
    Independent Fit 	&$16$		&$-1.4209$	\\
	Independent Fit     &$32$		&$-1.4074$	\\
	Constrained Fit
	\tablenote{$\:$Here we calculate $\Gamma_{4}$ using the parameters
	$u=0.228$ and $c=0.45$.}
             &$16$,$32$      &$-1.4168$\\

& \multicolumn{1}{c}{Simulations}\\
\tableline
	Tsypin\cite{T=T_c} &$16$		&$-1.424(3)$	\\
	Tsypin\cite{T=T_c} &$32$		&$-1.410(3)$	\\
	Barber, {\em et. al.}\cite{barber} &$16$ &$-1.4239(6)$	\\
	Barber, {\em et. al.}\cite{barber} &$32$ &$-1.4095(18)$	\\
  \end{tabular}
}
\end{table}

\section{Concluding remarks}
\label{sec:conc}

We arrive at the
conclusion that a low order scaling form
provides an excellent fit to the order parameter distribution both
quite near the three-dimensional Ising model critical point and
exactly at criticality.  Our approximant, while at the lowest
nontrivial order in the interdimensional $\epsilon$ expansion,
nonetheless includes full renormalization group flows for the relevant
variables and thus has critical point (hyper-)scaling built in.  The
scaling result that at criticality ${\cal H} \sim
|M|^{2d/(d-2+\eta)}$, where $2d/(d-2+\eta) \sim 5.7$ is in close
agreement with Tsypin's~\cite{T<T_c,T>T_c} polynomial fit.  That this
ought to be so follows from straightforward scaling arguments.

The fitting parameters used were the nonuniversal metrical factors for the
reduced temperature, $t$, the fourth order coupling constant $u$
and the
parameter $c$, which controls the effects of finite size on the
determination of the quantity $\ell^{*}$ through the relationship
Eq.  (\ref{cutoff}).  This is certainly fewer than allowed; the
$\phi^4$ model and the Ising model are presumably in the same
universality class, but are not identical.  We also took the coupling
$u$ to be constant.  However, there is no reason to believe that the
Ising model at criticality is at its fixed point for coupling
constants.  Hence, corrections to scaling should certainly be allowed,
which in practice would mean the renormalization group flow for the
fourth-order coupling $u(\ell)$ could be included.  We have chosen not
to do so in order to minimize the number of free parameters.

The prefactor used by Binder~\cite{binderagain} and Tsypin~\cite{T>T_c,T<T_c}
is discussed in Section~\ref{sec:prefactor} within the setting
of a renormalization group calculation.  While this prefactor has only
limited effect on the quality of fits in the high and low temperature
phases, it plays a crucial role at the critical point.  This issue is
discussed in more depth in the Appendix, where the influence of the
prefactor on the universal ratio $\Gamma_{4}$ is also explored.

The fact that a low-order-in-$\epsilon$ expression for the
renormalized free energy of the Ising system reproduces simulation
data to such a high degree of accuracy has encouraging implications
with respect to the derivation of an equation of state applicable to
an $O(1)$ system in the immediate vicinity of a critical point.  In
addition, there is every reason to be hopeful that low order
corrections to the zero loop expression utilized in this work will
allow for an even higher precision fit to data.

\section*{acknowledgements}
D.J. is grateful to the National Science Foundation for support
of this work under DMR92-17935. He also thanks
D.  Boyanovsky for his interest and helpful comments.  We are
especially grateful to Dr.~M.~Tsypin for bringing this problem to our
attention, for providing his simulation data and for useful comments
throughout the progress of this work.  W.L.  acknowledges useful
conversations with Professor R.  Cousins.  J.R.  acknowleges the
support of NASA and thanks C.  Borg and V.  Dohm for highly enlightening
comments.

\pagebreak
\begin{appendix}
\section{Effect of the prefactor on $\Gamma_{4}$}

The prefactor represents a finite size correction to the free energy.
To see that this is so, it suffices to exponentiate the prefactor. The
contribution of the prefactor to the free energy is nominally
independent of the size of the system, while the leading order terms scale as
the system's volume.  Clearly, in the thermodynamic limit, the
prefactor is swamped by those terms.  Nevertheless, the prefactor
cannot be ignored, even as the volume approaches infinity.  That this
is so is evident when one plots out the order parameter distribution
at the critical point.  Given Eq.  (\ref{effp}), neglect of the
prefactor leads to an order parameter distribution at the critical
point ($t=0$, $h=0$) that has the following form
\begin{equation}
P(M) \propto e^{-AL^{d}|M|^{2d/(d-2+\eta)}} \ \ .
\label{a1}
\end{equation}
Such a distribution does not manifest one important feature of the
critical point order parameter distribution, and that is a depression
in the in the neighborhood of $M=0$.  The absence of this feature has
an effect on the fit to data, and in addition, on the quantity
$\Gamma_{4}$, defined in Eq.  (\ref{Gamma}).  As an indication of the
importance of the prefactor on this universal ratio, we first
calculate its value in the absence of this contribution. We make use of
the general relationship
\begin{eqnarray}
\int_{0}^{\infty}M^{x}e^{-AM^{2d/(d-2+\eta)}} &=& A^{-(p+1)2d/(d-2+
\eta)}\frac{d-2+ \eta}{2d}\int_{0}^{\infty}y^{(p+1)(d-2+\eta)/2d -
1}e^{-y}dy \nonumber \\
&=& \frac{d-2 + \eta}{2d}A^{-(p+1)2d/(d-2+ \eta)}
\Gamma\left(\frac{(p+1)(d-2 + \eta)}{2d}\right)  \ \ ,
\label{a2}
\end{eqnarray}
where the function $\Gamma$ is the standard gamma function.  The
universal combination in Eq.~(\ref{Gamma}) is, then, equal to
\begin{equation}
\frac{\Gamma\left(\frac{5(d-2+\eta)}{2d}\right) \Gamma
\left(\frac{d-2+ \eta}{2d}\right)}{\Gamma
\left(\frac{3(d-2+\eta)}{2d}\right)} - 3  \ \ .
\label{a3}
\end{equation}
When $d=3$ and $\eta=0.037$, one finds
$\Gamma_{4} = -0.987593$.  As a comparison, the unrenormalized
Ginzburg-Landau theory predicts  $\Gamma_{4} = -0.81156$.
By contrast, numerical studies indicate
$\Gamma_{4} \simeq -1.4$.\cite{T=T_c,barber} While renormalization of the
theory changes $\Gamma_{4}$ in the right direction, the full scaling
form is inadequate to the challenge of reproducing the correct value of this
quantity.

The defect in the above analysis lies in the fact that finite size effects
have been ignored.  These effects are crucial at the critical point, where
the bulk correlation length is infinite.  As it turns out, the most
important finite size effect is the prefactor.  The influence of the
prefactor is dramatically highlighted when one ignores the finite size
contribution to Eq.~(\ref{cutoff}) for $\ell^{*}$.  If the coefficent
$c$ is set equal to zero, the one finds immediately
\begin{equation}
e^{\ell^{*}} = \left(3uM^{2}\right)^{-1/(d-2+\eta)} \ \ .
\label{a4}
\end{equation}
This means that the order parameter distribution is of the form
\begin{equation}
P(M) \propto M^{(2-\eta)/(d-2+\eta)} \exp \left[-\kappa
M^{2d/(d-2+\eta)} \right]
\label{a5}
\end{equation}
The quantity $\kappa$ in the above equation is a combination of the
size of the system, $L$, and the fourth order coupling constant, $u$.
This quantity scales out of the result for $\Gamma_{4}$ in the same way that
the constant $A$ divided out in the ratio in Eq.  (\ref{a3}).  The
result for $\Gamma_{4}$ following from Eq.~(\ref{a5}) is then
\begin{equation}
\Gamma_{4} = \frac{\Gamma\left(\frac{2-\eta +5(d-2+\eta)}{2d}\right) \Gamma
\left(\frac{2-\eta +(d-2+ \eta)}{2d}\right)}{\Gamma
\left(\frac{2 - \eta + 3(d-2+\eta)}{2d}\right)} - 3 \ \ .
\label{a6}
\end{equation}
Inserting appropriate values for $d$ and $\eta$ into Eq.~(\ref{a6}),
we obtain a $\Gamma_{4} = -1.69147$.  In utilizing the
prefactor as the {\em sole} finite size correction, we have overshot
the proper value of $\Gamma_{4}$.  Note, however, that the value obtained is
{\em independent of system size}.  Even though the
prefactor is a finite
size correction to the free energy, in the case of $\Gamma_{4}$ it is
fully as imporant as the ``leading order'' contributions to the order
parameter distribution.

As for the distribution $P(M)$, it is immediately evident that the
prefactor leads to a depression in the neighborhood of $M=0$.  In
fact, the order parameter distribution is forced to go to zero at
$M=0$ by the prefactor in Eq.~(\ref{a5}).  This is a more pronounced
effect than is desired.  The supression of the order parameter
distribution is reduced when when finite size effects are restored to
Eq.~(\ref{cutoff}).  It is the combination of these latter
finite size effects and the finite size correction embodied in the
prefactor that yields a proper order parameter distribution at the
critical point and a value of $\Gamma_{4}$ that is in agreement with
previous determinations of this quantity.

\end{appendix}

 \begin{figure}
 \centerline{\psfig{file=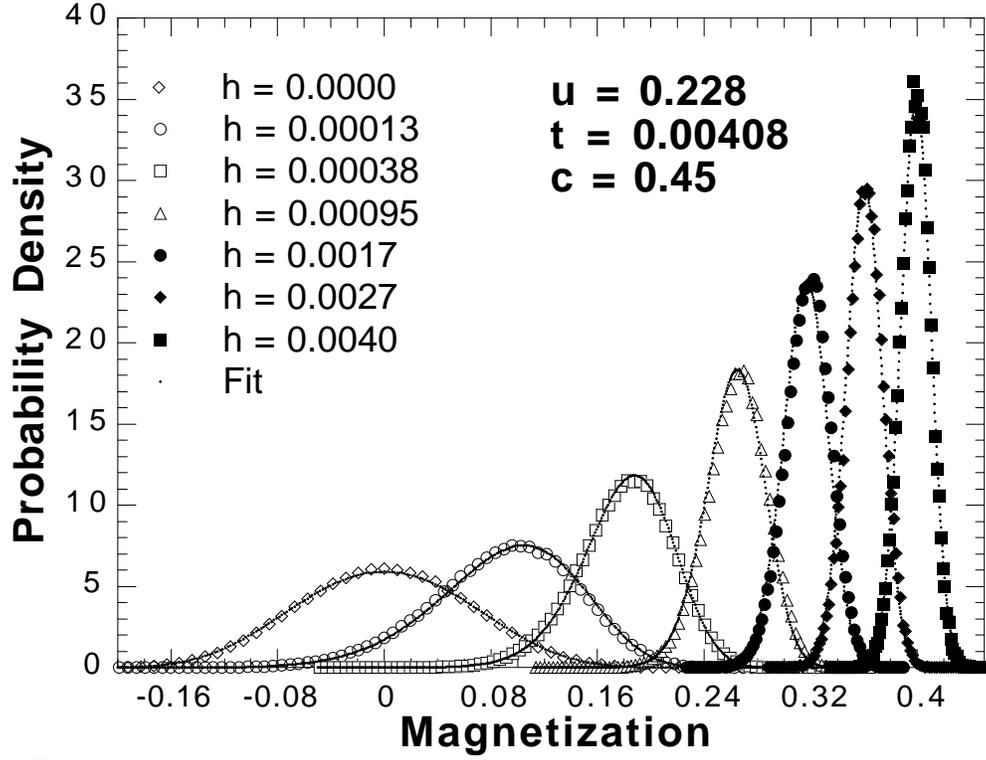,height=4in}}
 \caption{The order parameter
 distribution in the disordered phase, as obtained by Tsypin
 [1] (data points) and the results of the best constrained
 fit based on the effective Hamiltonian embodied in Eqs.  (\ref{Greffp})
 and (\ref{cutoff}).}
\label{fig1:disordered}
\end{figure}
\begin{figure}
\centerline{\psfig{file=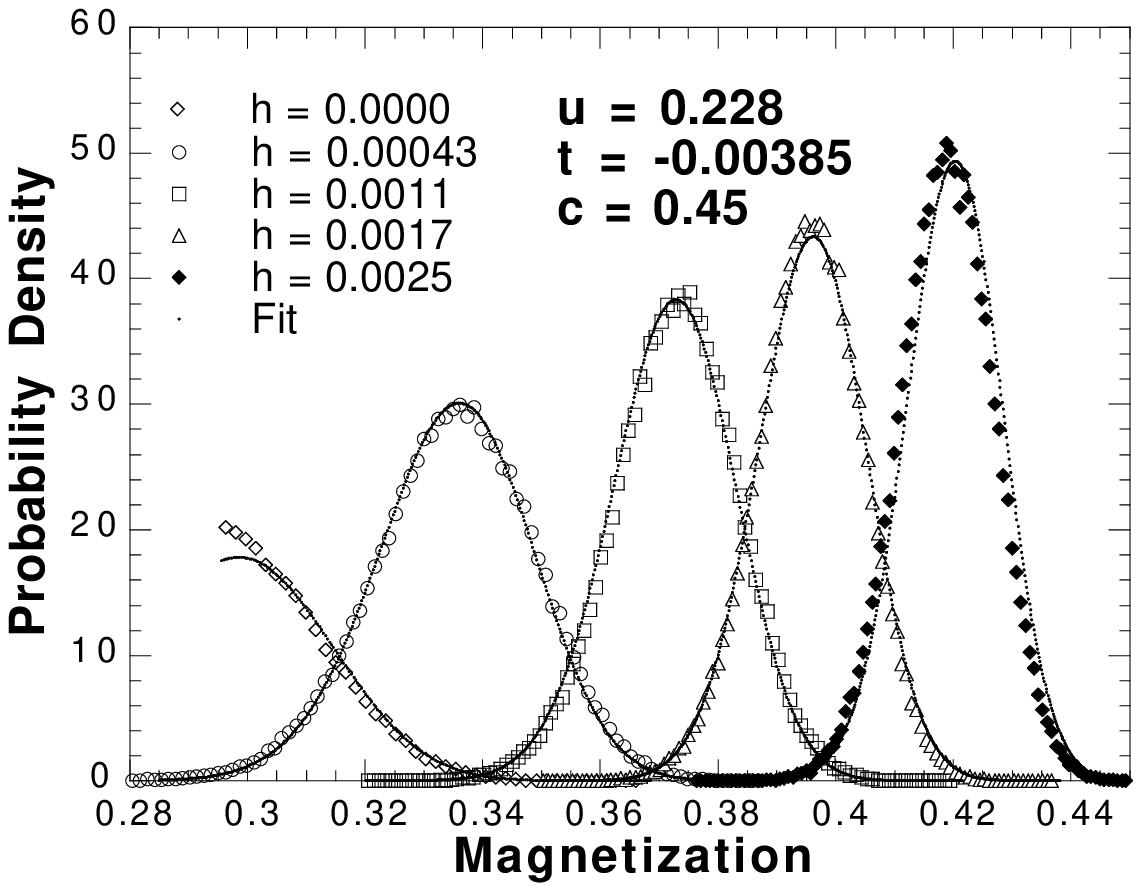,height=4in}}
 \caption{The order parameter distribution in the ordered phase, as obtained
by Tsypin[1] (data points) and the results of the best constrained fit based on
the effective Hamiltonian embodied in Eqs. (\ref{Greffp}) and (\ref{cutoff}).}
\label{fig2:ordered}
\end{figure}
\begin{figure}
\centerline{\psfig{file=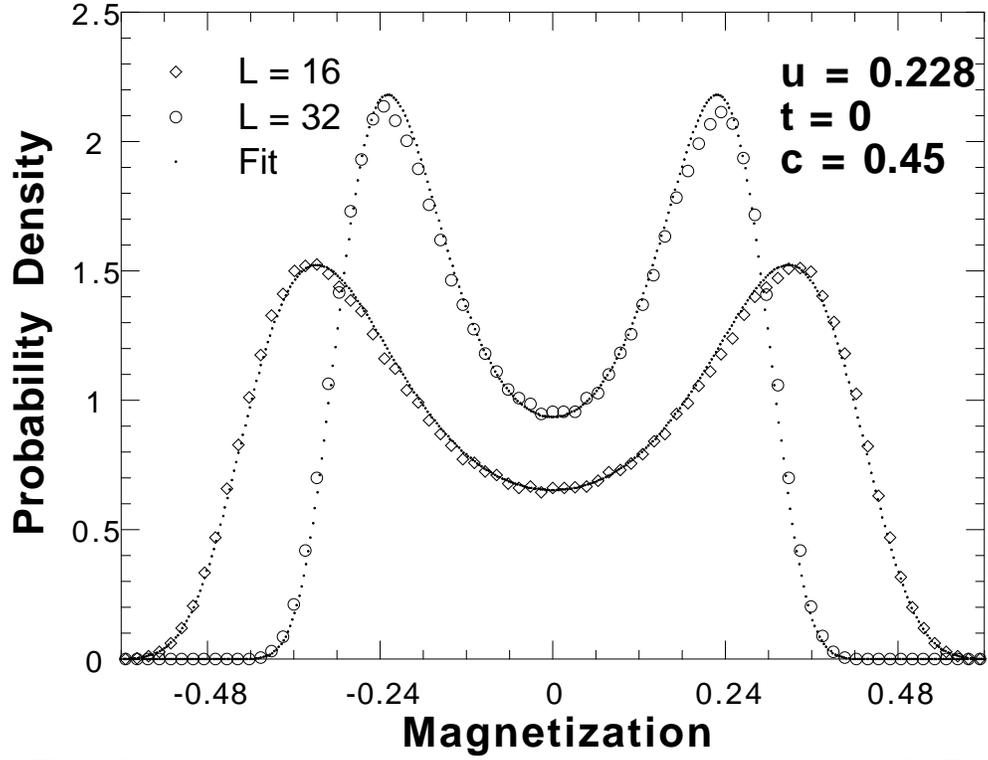,height=4in}}
\caption{The order parameter distribution of the system at $t=0$, as
obtained by Tsypin[3] and the results of the fit using Eqs.
(\ref{Greffp}) and (\ref{cutoff}). We include both the $16^3$ and $32^3$
systems
on the plot.}
\label{fig3:Critical}
\end{figure}


\begin{references}
\bibitem{chen} X.  S.  Chen and V.  Dohm, {\em
Physica A} {\bf 235}, 555 (1997).  We learned of this work, with which
there is substantial overlap in this article when the work reported
here was essentially complete.
\bibitem{refs} A comprehensive list of references can be
found in ref.  \cite{P&V}.
\end{references}
\end{document}